\documentclass{agujournal2019}
\usepackage{xspace}

\newcommand{\apj}{ApJ}
\newcommand{\apjl}{ApJ\xspace}

\newcommand{\apjs}{ApJS\xspace}

\newcommand{\aap}{A\&A}

\newcommand{\ssr}{Space Sci. Rev.}

\newcommand{\jgr}{J. Geophys. Res.\xspace}

\usepackage{url} %this package should fix any errors with URLs in refs.
\usepackage{lineno}
\usepackage{amsmath,amssymb}
\usepackage{natbib}
\usepackage[inline]{trackchanges} %for better track changes. finalnew option will compile document with changes incorporated.
\usepackage{soul}
\draftfalse

\journalname{JGR: Space Physics}
\newcommand{\td}[2]{\frac{d #1}{d #2}}

\newcommand{\pd}[2]{\frac{\partial#1}{\partial#2}}

\definecolor{ao}{rgb}{0.0, 0.5, 0.0}

\def\revision#1{{\color{black}{#1}}}

\begin{document}

%% ------------------------------------------------------------------------ %%
%  Title
%
% (A title should be specific, informative, and brief. Use
% abbreviations only if they are defined in the abstract. Titles that
% start with general keywords then specific terms are optimized in
% searches)
%
%% ------------------------------------------------------------------------ %%

\title{On the seed population of solar energetic particles in the inner heliosphere}

%% ------------------------------------------------------------------------ %%
%
%  AUTHORS AND AFFILIATIONS
%
%% ------------------------------------------------------------------------ %%

% Authors are individuals who have significantly contributed to the
% research and preparation of the article. Group authors are allowed, if
% each author in the group is separately identified in an appendix.)

% List authors by first name or initial followed by last name and
% separated by commas. Use \affil{} to number affiliations, and
% \thanks{} for author notes.
% Additional author notes should be indicated with \thanks{} (for
% example, for current addresses).

% Example: \authors{A. B. Author\affil{1}\thanks{Current address, Antartica}, B. C. Author\affil{2,3}, and D. E.
% Author\affil{3,4}\thanks{Also funded by Monsanto.}}

\authors{N. Wijsen\affil{1,2,3}, G. Li\affil{4}, Z. Ding\affil{1}, D. Lario\affil{2}, S. Poedts\affil{1,5}, 
R. J. Filwett\affil{6},
R. C. Allen\affil{7},
M. A. Dayeh\affil{8}
}

\affiliation{1}{Department of Mathematics/Centre for Mathematical Plasma Astrophysics, KU Leuven, Leuven, Belgium}
\affiliation{2}{NASA, Goddard Space Flight Center, Heliophysics Science Division, Greenbelt, MD 20771, USA}
\affiliation{3}{Department of Astronomy, University of Maryland, College Park, MD 20742, USA}
\affiliation{4}{Department of Space Science and Center for Space Plasma and Aeronomic Research, University of Alabama in Huntsville, Huntsville, AL 35899, USA}
\affiliation{5}{Institute of Physics, University of Maria Curie-Sk{\l}odowska, Lublin, Poland}
\affiliation{6}{Department of Physics and Astronomy, University of Iowa, Iowa City, IA 52242, USA}

\affiliation{7}{Johns Hopkins University Applied Physics Laboratory, Laurel, MD, 20723, USA}

\affiliation{8}{Southwest Research Institute, San Antonio, TX, 78238, USA}

%% Corresponding Author:
% Corresponding author mailing address and e-mail address:

% (include name and email addresses of the corresponding author.  More
% than one corresponding author is allowed in this LaTeX file and for
% publication; but only one corresponding author is allowed in our
% editorial system.)

% Example: \correspondingauthor{First and Last Name}{email@address.edu}

\correspondingauthor{Gang Li}{gangli.uahuntsville@gmail.com}

%% Keypoints, final entry on title page.

%  List up to three key points (at least one is required)
%  Key Points summarize the main points and conclusions of the article
%  Each must be 100 characters or less with no special characters or punctuation and must be complete sentences

% Example:
% \begin{keypoints}
% \item	List up to three key points (at least one is required)
% \item	Key Points summarize the main points and conclusions of the article
% \item	Each must be 100 characters or less with no special characters or punctuation and must be complete sentences
% \end{keypoints}

%% ------------------------------------------------------------------------ %%
%
%  ABSTRACT and PLAIN LANGUAGE SUMMARY
%
% A good Abstract will begin with a short description of the problem
% being addressed, briefly describe the new data or analyses, then
% briefly states the main conclusion(s) and how they are supported and
% uncertainties.

% The Plain Language Summary should be written for a broad audience,
% including journalists and the science-interested public, that will not have 
% a background in your field.
%
% A Plain Language Summary is required in GRL, JGR: Planets, JGR: Biogeosciences,
% JGR: Oceans, G-Cubed, Reviews of Geophysics, and JAMES.
% see http://sharingscience.agu.org/creating-plain-language-summary/)
%
%% ------------------------------------------------------------------------ %%

%% \begin{abstract} starts the second page

\begin{keypoints}

\item Seed particle populations in large SEP events may originate from flare remnants

\item  We examine the distribution of seed particles in the inner heliosphere by combining a magnetohydrodynamic (MHD) solar wind code and a  particle transport code

\item We obtain the longitudinal profile of the seed population at 1~au and show that the perpendicular diffusion coefficient is a key parameter 

\end{keypoints}

\begin{abstract}
Particles measured in large gradual solar energetic particle (SEP) events are believed to be predominantly accelerated at shocks driven by coronal mass ejections (CMEs).  Ion charge state and composition analyses suggest that the origin of the seed particle population for the mechanisms of particle acceleration at CME-driven shocks is not the bulk solar wind thermal material, but rather a suprathermal population present in the solar wind. 
This suprathermal population could result from remnant material accelerated in prior solar flares and/or preceding CME-driven shocks.     
In this work, we examine the distribution of this suprathermal particle population in the inner heliosphere by combining a magnetohydrodynamic (MHD) simulation of the solar wind and a Monte-Carlo simulation of particle acceleration and transport. 
Assuming that the seed particles are uniformly distributed near the Sun by solar flares of various magnitudes, we study the longitudinal distribution of the seed population at multiple heliocentric distances. We consider a non-uniform background solar wind, consisting of fast and slow streams that lead to compression and rarefaction regions within the solar wind. 
Our simulations show that the seed population at a particular location (e.g., 1~au) is strongly modulated by the underlying solar wind configuration.
Corotating interaction regions (CIRs) and merged interactions regions (MIRs) can strongly alter the energy spectra of the seed particle populations. 
In addition, cross-field diffusion plays an important role in mitigating strong  variations of the seed population in both space and energy. 
\end{abstract}

\begin{plainLanguageSummary}
Large solar energetic particle events are a major concern of Space Weather. During these events, particles are accelerated to very high energies at shock waves driven by coronal mass ejections (CMEs). 
While the CMEs in these events are often similar in morphology and speed, the intensity of the accelerated particles can differ significantly from one event to another. 
One possible reason for these large variations may be attributed to the variability of the underlying seed population that gets injected into the acceleration mechanism occurring at the CME-driven shock. 
To be efficiently accelerated at a CME-driven shock, particles need an initial speed larger than the speed of the shock wave.
For this reason, it is often assumed that the seed population originates from the suprathermal population, rather than the bulk solar wind. 
These suprathermal particles can be generated near the Sun by, e.g., micro and nano flares. 
In this work, we use the three-dimensional MHD model EUHFORIA and the particle transport and acceleration model PARADISE to examine how solar wind structures affect the distribution of these suprathermal particles at different heliocentric distances. 
We find that the variation of the solar wind speed at the inner boundary can lead to a significant longitudinal variation of the suprathermal population in the inner heliosphere. 
Our study may offer an explanation for the variation of particle intensity in large SEP events. 
\end{plainLanguageSummary}
\section{Introduction}

Solar energetic particles (SEPs) are a major concern of space weather. 
The energy of ions in large SEP events can reach up to $\sim$ GeV/nuc. 
These particles are believed to 
be accelerated either  near solar flares sites, often referred as ``impulsive'' events, or at shocks driven by coronal mass ejections (CMEs), often referred as ``gradual'' events \citep{Reames.etal97}. 
In gradual SEP events, particles are believed to be accelerated at the CME-driven shock front via the diffusive shock acceleration (DSA) mechanism. Since fast CMEs are presumably able to drive strong shocks with large Alfv\'{e}n Mach numbers, they are thought to be efficient particle accelerators. 
However, although most of the intense SEP events are associated with fast and wide CMEs \citep{Reames95,Kahler96,Gopalswamy.etal02,Cliver.etal04,Tylka.etal05,Kahler.Vourlidas13},  many of such CMEs do not lead to large SEP events \citep{Kahler96, lario2020}.  
Among different factors, such a large event variability could be due to variations in the seed population that feeds into the DSA mechanism. 
Earlier, \cite{Kahler.etal00} suggested that the presence of ambient energetic particles prior to the event may be a key factor in determining the size of the SEP event. 
In fact, one key assumption of the DSA mechanism is that the accelerated particles are injected at a speed larger than the shock speed and must therefore originate from a suprathermal population instead of the thermal population of the solar wind. 

Composition and charge state studies  \citep{Mason.etal99, Mason.etal00, Klecker.etal07, Mewaldt.etal12} have revealed that the seed population in large SEP events could be ambient coronal material or remnant material accelerated in solar flares and/or by preceding CME shocks.
% \citep{Klecker.etal07, Mewaldt.etal12} also suggest that particles accelerated at large SEP events could be relic flare material or energetic coronal materials that are possibly accelerated at preceding CMEs.  
Studies by \citet{Gopalswamy.etal04} found that large SEP events tend to occur during periods when multiple CMEs erupt from the same and nearby active regions. 
This led to the development of the ``twin-CME" scenario of large SEPs by \citet{Li.etal05,Li.etal12} \citep[see also][]{Ding.etal13}.  
In the ``twin CME" scenario, a  preceding CME erupts within $13$ hours of the main CME and provides both an enhanced seed population and an enhanced turbulence level, facilitating an efficient acceleration at the main CME-driven shock \citep{Li.etal12}. 
\citet{Li.etal12} suggested that a large seed population not only leads to a large intensity of energetic particles, but also determines the maximum particle energy that can be achieved at a shock. 
This is because the particle acceleration process is strongly tied to the amplification of the upstream waves and the wave power is proportional to the seed particles \citep{Zank.etal00, Rice.etal03, Li.etal03}. 
In the scenario proposed by \citet{Li.etal12}, the enhanced seed population was assumed to last only $13$ hours. 
However, recent work by \citet{Zhuang2021ApJ...921....6Z} showed that although the very large SEP events tend to have preceding CMEs within $\sim 13$ hours, many large SEPs do not have preceding CMEs within this time window. 
Therefore, if the seed population of large SEP events is due to preceding CMEs/shocks, it can exist in the inner heliosphere for an extended period of time ($>13 $ hours). 
It is therefore important to understand how this seed population is distributed in the inner heliosphere and how the underlying solar wind can influence its evolution. 

Solar flares tend to be more frequent than CMEs, suggesting that
the seed population for large SEP events could also be formed by remnant material accelerated in these more abundant solar flares \citep{Mason.etal99,Mason.etal00}. 
Recent observations from Solar Orbiter have shown numerous microflares in Extreme Ultraviolet  images
\citep{berghmans_2021} and in X-rays \citep{Saqri_2022}, even in the quiescent solar corona.
Magnetic reconnection seems to play a critical role in these abundant microflares \citep[e.g.,][]{Chen_Y_2021,Hou_Z_2021, Li_ZF_2022}.
Therefore, we argue that the particle acceleration occuring in these small flares observed even during solar minimum \citep[e.g.,][]{Christe2008ApJ...677.1385C}, may fill the inner heliosphere with suprathermal particles.
That is, we suggest that these small-scale flares can provide an ever-present seed population.
This seed population is generated close to the Sun and will propagate into the solar wind. 
Because they are of higher energy compared to the background solar wind plasma, they propagate mostly along the interplanetary magnetic field (IMF) and will experience the effects produced by  rarefaction and compression regions of the IMF resulting from inhomogeneities of the  solar wind. Acceleration by compressive waves or shocks near corotating interaction regions (CIRs) may also exist and affect the distribution of these seed particles in the inner heliosphere. 
Clearly, it is important to understand how various solar wind structures can affect this seed population.
%\gangli{== Possible future project: investigating the seed population variation as a function of solar cycle. Comparing the observations with model to constrain the flare acceleration process.  Both the frequency of micro flares, and the maximum energy will enter into the simulation. In our model, they represent the absolute intensity, the spectral index and the maximum energy....  ==}
%which although is solar-cycle dependent, but still noticeable during solar minimum. 

Suprathermal seed populations may be also accelerated in-situ in the solar wind from the thermal pool by compressional turbulence, as advocated by \citet{Fisk2006ApJ...640L..79F,Fisk2008ApJ...686.1466F}.
This mechanism predicts an ubiquitous spectral tail in the form of $j (E) \sim E^{-1.5}$
(or $f(v) \sim v^{-5}$) in the energy range of a few keV up to tens of keV. However, at above $50$ keV,  
in-situ observations show a wide range of spectral indices varying 
between $1.2$ and $2.9$ \citep[e.g.,][]{Mewaldt07,Dayeh_2009ApJ...693.1588D,Dayeh2017ApJ...835..155D, Desai2010, Filwett2019}.
Another observational signature of the seed population concerns its longitudinal variation. Multiple spacecraft observations from STEREO-A, B, and ACE have revealed that the seed population intensity can vary largely as a function of longitude. 

%\gangli{Maher: we can refer to your recent observational paper to be submitted on this matter.}

%Clues about the origin of superthermal (?)
%Composition of suprathermal, not of solar wind origin, but flare remnant...  [Mason et al...]
What is clear from the above discussion is that the origin and evolution of the suprathermal seed population in the inner helisophere is still a question of debate.  
In this work, we use numerical simulations to examine the seed population distribution in the inner heliosphere. 
We adopt the scenario where the origin of the seed particles occurs in micro and nano flares, and assume that they are  uniformly distributed at the inner boundary of our simulation domain, which is located at 0.1~au from the Sun (see details below). 
We then study the transport of this seed population in a non-uniform solar wind which is composed of multiple fast and slow wind streams. 
The non-uniform solar wind is modelled using the EUropean Heliospheric FORecasting Infromation Asset (EUHFORIA) model \citep{pomoell2018} and the particle transport is followed using the PArticle Radiation Asset Directed at Interplanetary Space Exploration (PARADISE) model \citep{wijsenPHD2020}. 
We find that velocity shears in the solar wind can affect the spatial distribution of the seed particles, leading to a significant longitudinal variation of these seed particles in the inner heliosphere. Furthermore, cross-field diffusion plays an important role in mitigating strong variations of the seed population in both the configuration space and the energy space. 

%Radial variation and the spectral features of the seed population at different locations are also examined.

The paper is organized as the follows: in section \ref{sec:Model} we describe the two models, EUHFORIA and PARADISE used in our work; in section \ref{sec:Results}, we present our simulation results; we end with a discussion in section ~\ref{sec:conclusions}.

\section{Model Description}\label{sec:Model}
%We use EUPHOREA
\begin{figure}[ht]
    \centering
   \includegraphics[width=0.75\textwidth]{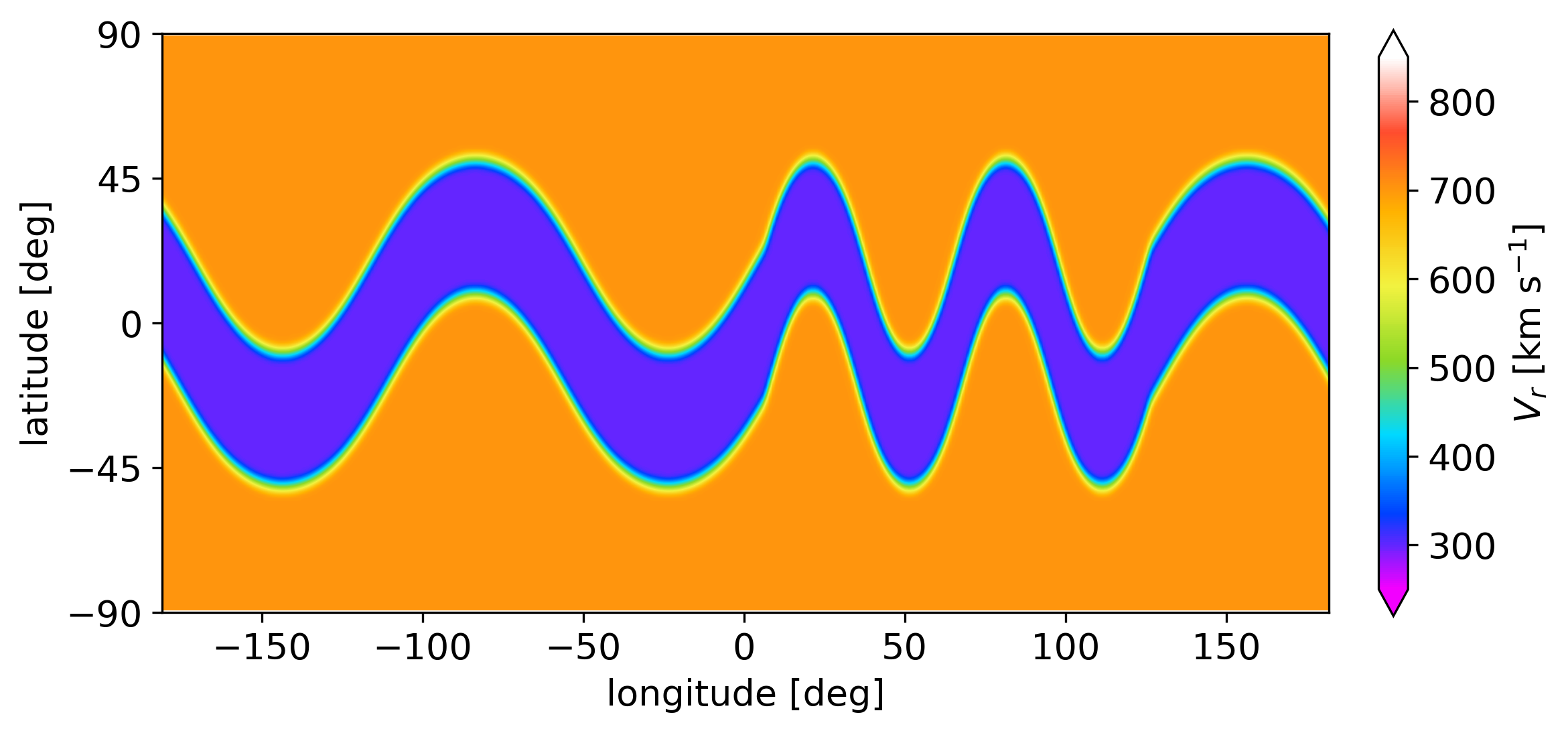}
    \caption{The radial speed profile at the inner boundary of the EUHFORIA simulation.} \label{fig:BC}
\end{figure}
\begin{figure}[ht]
    \centering
   \includegraphics[width=0.99\textwidth]{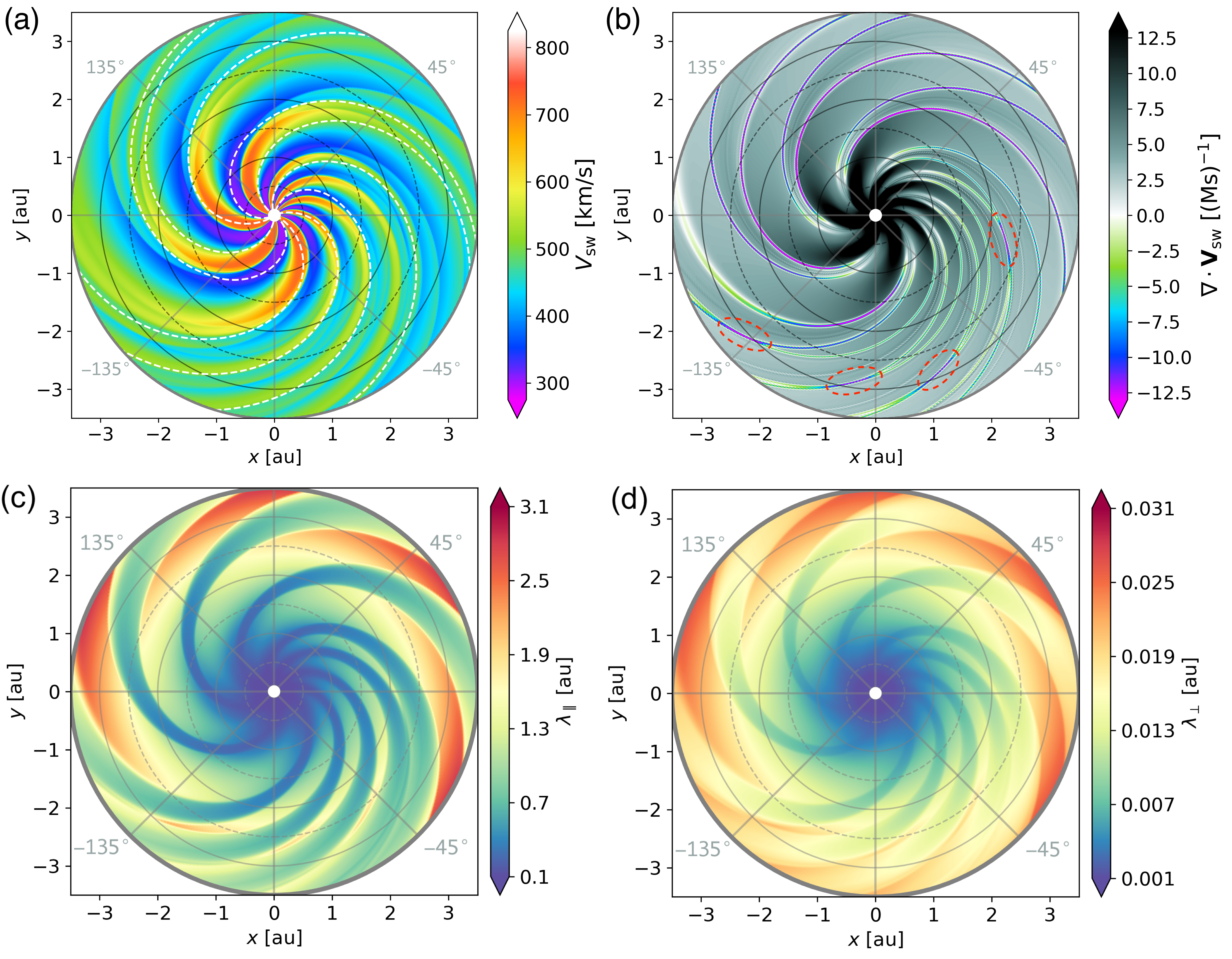}
    \caption{The EUHFORIA solar wind and the assumed parallel and perpendicular mean free paths in the solar equatorial plane.  
    Panel (a) shows the  speed of the solar wind.  
    White dashed lines are IMF lines projected on the solar equatorial plane. 
    Panel (b) show the divergence of the solar wind velocity, $\nabla\cdot\vec{V}_{\rm sw}$. The compression waves bounding the various CIRs can be recognized as the colorful spiral-shaped structures. The dashed red ovals indicate the mergings and crossings of forward and reverse compression waves.  
    Panels (c) and (d) show, respectively, the parallel and perpendicular mean free paths for $10$ keV protons in the PARADISE simulations.
    } \label{fig:MDI}
\end{figure}

To study the evolution of the seed particle population in the inner heliosphere, we use the PARADISE model \citep{wijsenPHD2020} which 
%simulates the transport and acceleration of energetic particles in the heliosphere. This model 
evolves energetic particle distributions through a background solar wind generated by the three-dimensional MHD model EUHFORIA \citep{pomoell2018}.
In this study, EUHFORIA is used to generate a synthetic solar wind 
which contains various fast and slow wind streams of different size.  Note that our synthetic solar wind is not intended to reproduce a realistic solar wind configuration generated during one particular Carrington rotation, but should instead be view as containing an ensemble of different solar wind structures that may be encountered by the seed population when it propagates through the inner heliosphere.

To generate our synthetic wind, the bimodal speed profile shown in Figure~\ref{fig:BC} at the inner boundary of the model, located at  $R_b=0.1$~au.  Following the standard set-up of EUHFORIA \citep{pomoell2018}, the thermal pressure at the inner boundary is chosen to be constant and equal to $P = 3.3$~nPa and the number density is prescribed as $n = n_{\text{fw}} (v_{\text{fw}} / v_r)^2$, with $v_{\text{fw}}= 700$~km~s$^{-1}$ and $n_{\text{fw}} = 300$~cm$^{-3}$.  
This choice ensures that the solar wind has a constant kinetic energy density at $R_b=0.1$~au.  The resulting simulated solar wind at 1~au attains speeds between 317~km/s and 722~km/s and the number densities varies between $1.6$ cm$^{-3}$ and $39$ cm$^{-3}$. At the inner boundary of the simulation, the radial component of the magnetic field ($B_r$) is assumed to be positive and has a value of 300~nT. The value of the azimuthal component ($B_\phi$) is chosen such that the electric field is zero in the corotating frame. This is necessary to achieve a steady-state solution in that frame.
\revision{It is worth noting that the monopolar nature of $B_r$ implies that we do not consider heliospheric current sheet (HCS) in our simulation. We point out that at the HCS, the reversal of magnetic field direction can cause energetic SEPs and galactic cosmic rays (GCRs) to drift along the HCS \citep[e.g.,][]{Jokipiilevyhubbard1977,KotaJokipii1983,Dalla_etal_2020}. However, for the low-energy protons studied in this work, HCS drifts are expected to be small \citep[e.g.,][]{battarbee2018}, so we have neglect them. Additionally, while the HCS can be included in a EUHFORIA simulation \citep[e.g.,][]{scolini2021}, the modeled HCS is typically wider than the observed HCS due to limited spatial resolution of the simulation. This can cause non-physical effects on the suprathermal particles modeled by PARADISE. In future work, this issue will be addressed by using adaptive mesh refinement (AMR) to increase the resolution of the computational mesh at the current sheet \citep[e.g.,][]{baratashvili22}.}

Figure~\ref{fig:MDI}a shows the solar wind speed $V_{\rm sw}$  in the solar equatorial plane. Fast and slow solar wind streams with different longitudinal extends can be clearly spotted.
Corotating interaction regions (CIRs) form where a slow solar wind stream is followed by a faster solar wind stream. Rarefaction regions form in the tails of  fast solar wind streams, before the arrival of the following slow solar wind.    
These structures can also be seen in Figure~\ref{fig:MDI}b, which shows $\nabla\cdot\vec{V}_\text{sw}$ in the solar equatorial plane. 
The gray shaded regions correspond to the locations where the solar wind is expanding, with the fast solar wind streams and their trailing rarefaction having the darkest colors. Compression regions can be identified by the colorful (yellowish and blueish) spiral shapes bounding the CIRs. We associate these compression regions, that appear in pairs, with the forward and reverse compression waves typically observed preceding and trailing the CIRs. These compression waves can steepen into reverse and forward shocks further away from the Sun.  Reverse shocks can be seen to form beyond ${\sim}1.5$ au in our simulation. See below for more discussions. 
The red-dashed ovals in Fig.~\ref{fig:MDI}b indicate the locations where the forward and reverse shocks of two neighboring CIRs cross each other. Further out and beyond this shock crossing, the two CIRs can be said to form a merged interaction region \citep[MIR; e.g., ][]{Dryer1976,Dryer1978, WhangBurlaga1985, Burlaga_etal1990}. 
We remark that in the simulation it are only the smaller HSSs that produce MIRs within $\lesssim 3$~au, as the CIRs produced by these HSSs are more closely spaced in longitude. In general, the formation location of MIRs will strongly depend on the extend and separation of the HSSs driving the merging CIRs and also on the local plasma conditions as they determine the propagation speeds of the reverse and forward shocks. 

%As discussed below, the suprathermal particle seed population might be  significantly different near CIRs, MIRs, and rarefaction regions, as these regions allow for changes in the particle acceleration or deceleration efficiency.

\revision{In this study, the Monte-Carlo model PARADISE is utilised to evolve a suprathermal particle seed population through the EUHFORIA solar wind domain. This is accomplished by solving the focused transport equation \citep[see][for a recent review]{vandenBerg2020}
\begin{align}
\pd{f}{t} 
+\left(\vec{V}_{\rm sw}+ \mu v\vec{b}\right)\cdot\nabla{f}
+
\left( \frac{1-3\mu^2}{2}(\vec{b}\vec{b}:\nabla\vec{V}_{\rm sw}) - \frac{1-\mu^2}{2}\nabla\cdot\vec{V}_{\rm sw} -\frac{\mu }{v}\vec{b}\cdot\td{\vec{V}_{\rm sw}}{t}\right) p \pd{f}{p} 
\notag\\+
\frac{1-\mu^2}{2}\left(v \nabla\cdot\vec{b} + \mu \nabla\cdot\vec{V}_{\rm sw} - 3 \mu \vec{b}\vec{b}:\nabla\vec{V}_{\rm sw}- \frac{2}{v}\vec{b}\cdot\td{\vec{V}_{\rm sw}}{t} \right)\pd{f}{\mu}
+
\td{p}{t} 
= \notag\\
\pd{}{\mu}\left(D_{\mu\mu}\pd{f}{\mu}\right) + \nabla\cdot\left(\boldsymbol{D}_\perp\cdot\nabla f\right).\label{eq:fte}
\end{align}
In this equation, $f(\Vec{x}, p,\mu, t)$ denotes the gyro-averaged particle distribution function, which depends on the time $t$,  the  spatial coordinate $\vec{x}$, the momentum magnitude ${p}$,  and the pitch-angle cosine $\mu$. Moreover, the  differential particle flux, $\vec{j}(E)$, is related to $f$ through $\vec{j}(E) = p^2f(p)$.
The other terms appearing in Equation~\eqref{eq:fte} are the particle speed $v$, the solar wind velocity $\vec{V}_{\rm sw}$, and the unit vector   $\vec{b}$, which points in the direction of the ambient IMF. In the PARADISE model, $\vec{V}_{\rm sw}$ and $\Vec{b}$ are obtained from the EUHFORIA model.  }
\revision{
The two terms appear on the right hand side of Equation~\eqref{eq:fte} describe diffusion. It includes a pitch-angle diffusion process, described by $D_{\mu\mu}$, and a spatial diffusion process perpendicular to the background magnetic field, which is described by the tensor $\boldsymbol{D}_\perp = D_\perp (I- \Vec{b}\Vec{b})$, with $I$ denoting the identity matrix.}  These diffusion processes model the effect of a fluctuating magnetic field $\delta \vec{B}$ on the particle transport. 
We assume that this magnetic turbulence is magnetostatic, incompressible, and can be decomposed into a slab and a 2D component \citep[e.g.,][]{Matthaeus1990JGR....9520673M}. 
We further assume that the ratio of the magnetic variance to the background magnetic field magnitude scales as

\begin{equation} \label{eq:varB_radialDependence}
  \frac{\delta B^2}{B_0^2} =
  \begin{cases}
  \Lambda_0 \left(r/r_0\right)^{\alpha_1} &  r  \leq r_1 = 0.5\text{ au}\\ 
  \Lambda_1  \left(r/r_0\right)^{\alpha_2} & r_1 < r  \leq r_2 = 2.0 \text{ au} \\ 
  \Lambda_2   & r_2 < r
  \end{cases}   
\end{equation}
where $B_0$ is the background IMF, $r_0 = 0.1$~au, 
$\Lambda_0 = 0.1$, 
$\Lambda_1 = \Lambda_0 (r_1/r_0)^{\alpha_1-\alpha_2} \approx 0.15$,
$\Lambda_2 = \Lambda_1 (r_2/r_0)^{\alpha_2} \approx 0.32$,  
with $\alpha_1 = 0.5$,
and $\alpha_2 = 0.25$. 
The choice of  equation~\eqref{eq:varB_radialDependence} is to capture the  different radial dependence of the background magnetic field and the turbulent field $\delta B$.  Within $0.5$~au, the magnetic field is largely radial and scales on average as $\sim r^{-1.5}$; beyond $2$ au, the magnetic field is largely azimuthal and scales on average $\sim r^{-1}$. Recent observations by Parker Solar Probe (PSP) \citep{Adhikari2020} on the turbulent $\delta B^2$ suggests that it has a $\sim r^{-2.54}$ dependence within $1$ au, which is somewhat shallower than the WKB approximation of $r^{-3}$. Beyond $2$ au, it also hardens and we assume its ratio to $B_0^2$ is $r$-independent.
Following \citet{bieber1994}, we assume that $\delta B^2_{\rm slab} = 0.2\delta B^2$ and $\delta B^2_{\rm 2D} = 0.8\delta B^2$. 
The 2D and slab  correlation length are prescribed as $\ell_{2D} = (0.0074~\text{au}) (r / 1~\text{au})^{1.1}$ and $\ell_{\text{slab}} = 3.9\times\ell_{2D}$, respectively  \citep{strauss2017,weygand2011}. 

The functional form of the pitch-angle diffusion coefficient is derived from quasi-linear theory (QLT), and scaled such that the parallel mean free path equals \citep{teufel2003}
\begin{equation}
\label{eq:LambdaParl}
\lambda_{\parallel} = \frac{3 s}{\pi (s - 1)} \ell_{\rm slab} R^2 \frac{B_0^2}{\delta B_{\rm sl}^2} \left[ \frac{1}{4} + \frac{2 R^{-s}}{(2 - s)(4 - s)} \right],
\end{equation}
where $s$ is the spectral index of the turbulence power spectrum and $R = R_L/\ell_{\rm slab}$, with $R_L$ the Larmor radius for a $90^{\circ}$ pitch angle. 
Figure~\ref{fig:MDI}c shows the resulting parallel mean free path in the solar equatorial plane for $10$~keV protons. 
The strong dependence of $\lambda_\parallel$ on the type of the underlying solar wind is a consequence of the particles' gyro-frequency, and hence the wave-particle resonance condition dependence on the background magnetic field.
In particular, the figure illustrates that in the simulation, the rarefaction region is characterised by a large parallel mean free path.
This  reflects the findings of, for example, \citet{McDonald1985} and  \citet{Desai2020}, that particle scattering (and hence also adiabatic deceleration) can be strongly reduced inside rarefaction regions. 
The latter may be a consequence of the decay of the amplitude in Alfv\'enic fluctuations that is seen in rarefaction regions \citep[e.g.,][]{carnevale2022}. 

The cross-field diffusion coefficient used in the PARADISE simulations is based on the nonlinear guiding centre (NLGC) theory of \citet{matthaeus2003}, modified by \citet{engelbrecht2019} to include an explicit dependence on the particle's pitch angle:
\begin{equation}\label{eq:nlgc}
    D_\perp =\mu^2 v \lambda_\parallel^{1/3}\left( a^2\sqrt{3\pi}\frac{2\nu -1 }{\nu} \frac{\Gamma(\nu)}{\Gamma(\nu -1/2)} \frac{\delta B^2_{2D}}{B_0^2} \ell_{2D} \right)^{2/3},
\end{equation}
where $v$ is the particle speed, $\mu$ the cosine of the pitch angle, $\nu = s/2$, $\Gamma$ denotes the gamma function, and $a$ is a free parameter %\gangli{introduced by \citet{Shalchi2010},}
which we set equal to $\sqrt{1/3}$, consistent with the work of \citet{matthaeus2003}.
Figure~\ref{fig:MDI}d shows the resulting perpendicular mean free path $\lambda_\perp = \frac{3}{2v} \int_{-1}^{1} D_\perp(\mu) d\mu$ in the solar equatorial plane for $10$~keV protons.
The azimuthal variation of the perpendicular mean free path is due to its dependence on $\lambda_\parallel^{1/3}$. The ratio $\lambda_\perp/\lambda_\parallel$ ranges from $10^{-3}$ to $3 \times 10^{-2}$, with the minimum and maximum values obtained in the rarefaction regions and CIRs, respectively. 

We assume a steady and uniform distribution of particles at the inner boundary $R_b=0.1$~au representing a population of particles remnant from e.g., multiple microflare injections.  This assumption allows us to investigate the effect of solar wind shear on the propagation of these particles throughout  the inner heliosphere. 
We do not consider the propagation of particles from near the photosphere (PS) where they were supposedly injected
to the source surface in this work. 
The coronal magnetic field is 
highly dynamic and irregular. At the PS, flares are not uniformly distributed, but presumably,
the overall effect of such a field is to smear out the spatial distribution of the flare source. 

The differential intensity injected at the inner boundary is assumed to be of the functional form 
\begin{equation}\label{eq:inj}
    j_{\rm inj }(E) = j_0E^{\gamma}\exp(-E/E_0),
\end{equation}
and protons from 10~keV up to 26 MeV are injected in the PARADISE model. By way of example, we choose the roll-over energy $E_0 = 500$~keV and the spectral index $\gamma = -2.5$.  
This is softer than the $\gamma = {-1.5}$ predicted by the compressional wave acceleration models \citep{Fisk2006ApJ...640L..79F}.
Moreover, we assume that $j_{\rm inj }$ is independent of time. 
 This is accomplished by first calculating the Green's function solution of the focused transport equation obtained by injecting the particles all at time $t=0$. We then obtain the steady state solution for a time-constant particle injection by convolving the Green's function solution  at different times with a time-independent function. 
 %We are particularly interested in the longitudinal variation of the seed particle intensity at different energies and how  the seed particle spectrum varies with longitude.
\revision{The constant $j_0$ in Eq.~\eqref{eq:inj} is chosen so that the simulated 10~keV protons have an average differential intensity of $10^3$~(cm$^2$~s~sr~MeV)~$^{-1}$ in the regions at 1~au where the EUHFORIA solar wind speed drops below 320~km/s. This normalization is based on the results of \citet{gloecker2008}, who conducted an ensemble average of differential proton intensities measured by ACE during quiet times in 2007. These selected quiet time periods were defined as having a solar wind speed below 320~km/s.  We would like to point out that the constant $j_0$ for all the presented simulations is obtained from the simulation without cross-field diffusion. This is done to facilitate a meaningful comparison between the different simulations. 
}

\section{Results}\label{Results} \label{sec:Results}

\subsection{Spatial distribution of the seed population within $3.5$ au}

\begin{figure}[ht]
    \centering
    \includegraphics[width=0.99\textwidth]{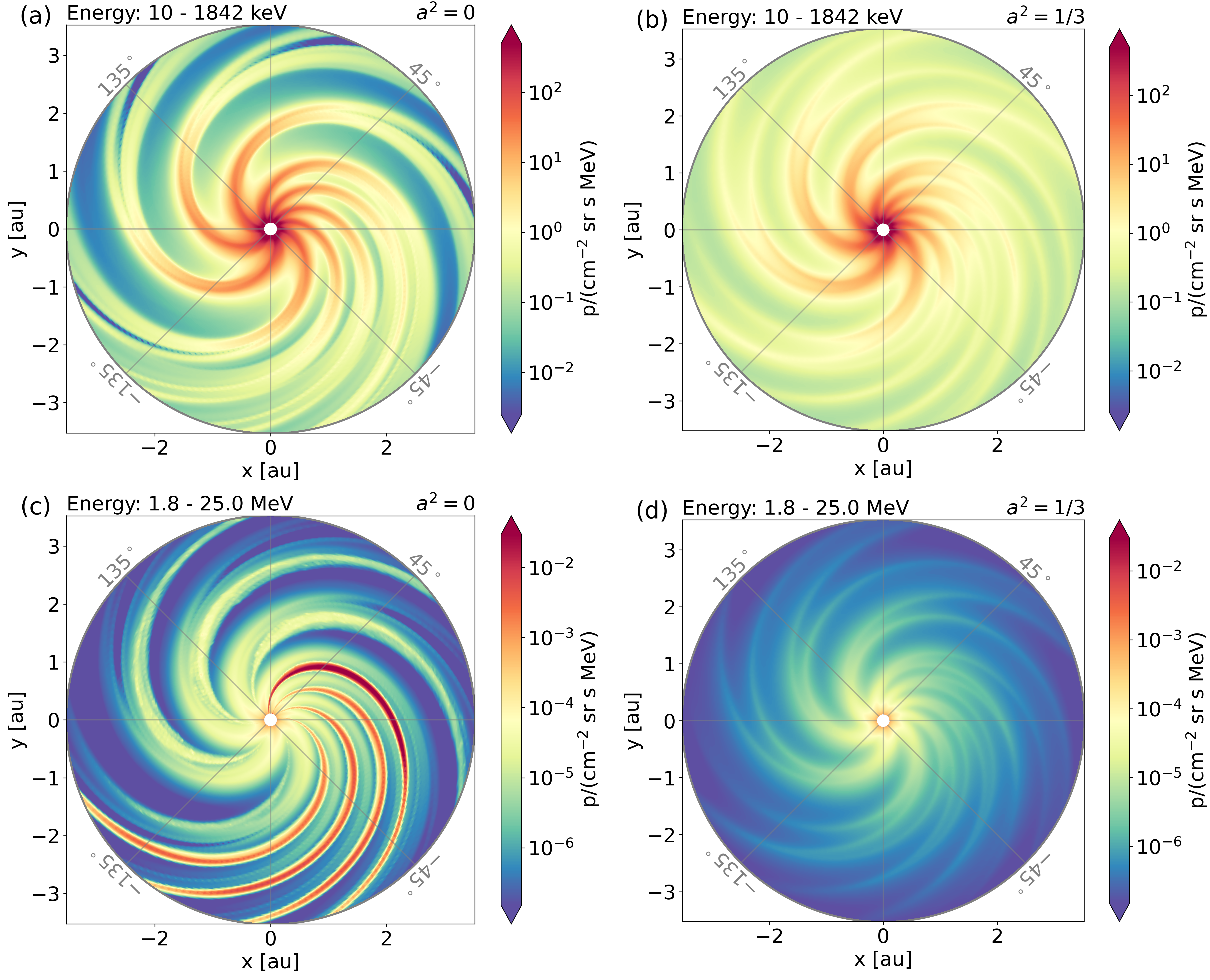}
    \caption{Omnidirectional particle intensities in the solar equatorial plane for the energy channels $10$ keV -- $1.8$ MeV (top row) and $1.8$ MeV -- $25$ MeV (bottom row). 
    The left column corresponds to simulations without cross-field diffusion ($a^2=0$), and the right column to simulations with cross-field diffusion with $a^2=1/3$. } \label{fig:EPs_in_eqplane}
\end{figure}

Figure~\ref{fig:EPs_in_eqplane} shows the simulated omnidirectional particle intensities for two energy ranges: $10$ keV -- $1.8$ MeV in the top row and $1.8$ MeV -- $25$ MeV in the bottom row. 
The left and right columns show the results for the simulations without and with cross-field diffusion, respectively (i.e. $a$=0 and $a$=$\sqrt{1/3}$ in Eq.\ref{eq:nlgc}, respectively). Comparing (b) to (a) and (d) to (c), we can see that the simulated intensity without cross-field diffusion shows a strong longitudinal variation.
For the lower energy channel, this variation reflects the underlying solar wind configuration. 
That is, comparing Figure~\ref{fig:EPs_in_eqplane}a with Figure~\ref{fig:MDI}a and Figure~\ref{fig:MDI}b we see that the highest intensities are obtained inside the CIRs and the lowest intensities are obtained inside the rarefaction regions. 
This is of course because, without cross-field diffusion,  suprathermal particles can only propagate along the background magnetic field lines, which are frozen into the solar wind. As a result these suprathermal particles will feel the compression and rarefactions of the underlying solar wind. 
When cross field diffusion is included (Figure~\ref{fig:EPs_in_eqplane}b), the gradients in the particle intensity between neighboring field lines will lead to a current, which tends to homogenize particle distribution across these field lines. This is illustrated in Panel~b of Figure~\ref{fig:EPs_in_eqplane}. % illustrates that, as expected, cross-field diffusion tends to isotropize the particle distributions.

The behaviour of the high energy channel $1.8$ MeV -- $25$ MeV  is significantly different from the lower energy channel. 
For the higher energy channel, Figure~\ref{fig:EPs_in_eqplane}c shows that the largest intensities  are obtained along background magnetic field lines that are magnetically connected to the forward-reverse shock crossings as indicated by the red ovals in Figure~\ref{fig:MDI}b. 
This is the result of an efficient ongoing interplanetary particle acceleration process, as explained below.

In general, particle acceleration may occur in regions with converging solar wind flows, that is, in regions where $\nabla\cdot\vec{V}_\text{sw} < 0$ \citep[e.g.,][]{giacalone2002,leRoux2009,wijsen2019a}. 
In these regions, particles can gain energy by interacting with converging scattering centres (which are assumed to be frozen into the background solar wind). A particle in the simulation  will therefore be accelerated  when crossing the compression/shock waves bounding the CIRs. 
When multiple crossings occur, particles undergo a first order Fermi acceleration process, or diffusive shock acceleration (DSA) in the case when shocks are formed \citep{axford77,bell78}. 
For DSA to be efficient, elevated turbulence levels in the upstream and downstream regions are beneficial since a more turbulent medium leads to a smaller particle diffusion  coefficient $\kappa$ and a shorter acceleration time scale, which is proportional to $\sim \kappa/u^2$ with $u$ the background flow speed. Roughly speaking, stronger turbulence will increase the number of crossings of a particle at the shock front. 

In our simulations, a simple DSA scenario is not expected to be very efficient due to the relatively large mean free path of the particles across the compression waves (see Figure~\ref{fig:MDI}c). However, the simulation contains a scenario which ensures a very interesting variant of the DSA mechanism to work resulting from the formation of a region where the forward and reverse compression waves of the adjacent CIRs converge. We name this scenario a ``macroscopic collapsing magnetic trap" (MCMT).    
%other solar wind structures where particles can undergo an efficient acceleration mechanism, namely the regions where the 
The forward and reverse compression waves propagate in different directions in the solar wind frame and can eventually cross each other, forming an MIR \citep[e.g.,][]{Dryer1976,Dryer1978, WhangBurlaga1985, Burlaga_etal1990}.  
The configuration of the field line between these converging shock waves resembles that of a collapsing magnetic trap and lead to an efficient first order Fermi acceleration. Indeed,  as shown in Figure~\ref{fig:EPs_in_eqplane}c, high intensity regions for the high energy channel can be identified to  be along those field lines that connected to MIRs. 
In particular, the MIR that forms at $r \sim 2.25$~au in the simulation is seen to be the most efficient particle accelerator (corresponding to the topmost red stripe).

Figures~\ref{fig:EPs_in_eqplane}c ($a^{2}$=0)  and~\ref{fig:EPs_in_eqplane}d ($a^{2}$=1/3) allow us to compare the effects of cross-field diffusion in the distribution of 1.8-25 MeV particles.
The inclusion of cross-field diffusion has a significant effect on the particle acceleration process and the (longitudinal) distribution of accelerated particles.
The clear association between enhanced particle intensities and MIRs 
fades away. 
This is because the reverse-forward shock crossings constitute a relatively small fraction in longitudes, and once cross-field diffusion is included, particles have a high probability to be quickly removed from field lines that connect 
to these converging shock structures. 
Therefore, cross-field diffusion causes particles to escape the MCMT before they are accelerated significantly.

\subsection{Longitudinal variation of particle intensity} %{Intensity at different longitudes}
\begin{figure}[ht]
    \centering
   \includegraphics[width=\textwidth]{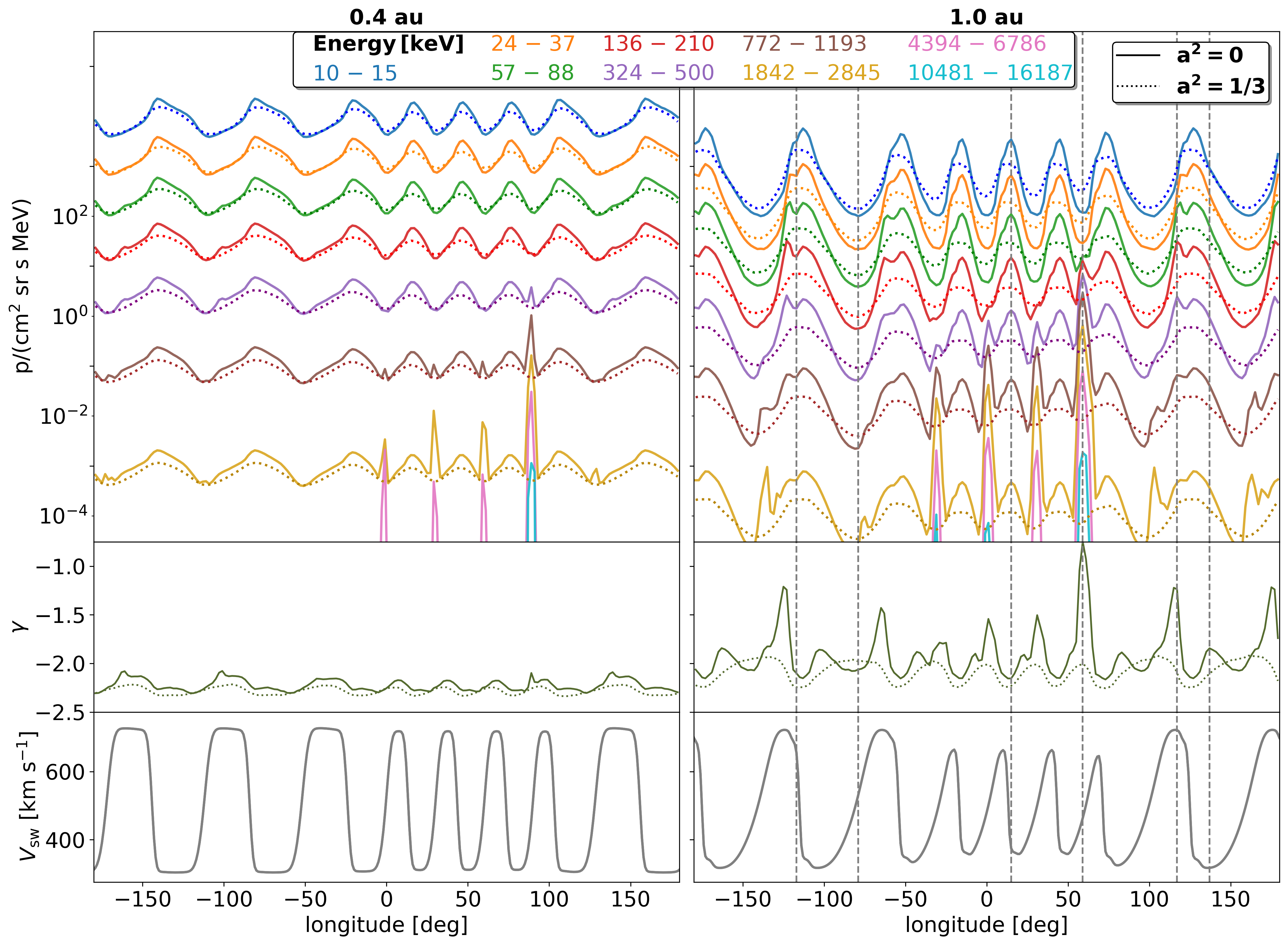}
    \caption{Longitudinal profiles of the simulations at $r=0.4$~au (left column) and $r=1.0$~au (right column) in the solar equatorial plane. The top two rows show the  omnidirectional intensities for different energy channels and for the different cross-field diffusion conditions, as indicated by the parameter $a^2$ in the panels' inset. The middle row gives the spectral index $\gamma$, obtained from fitting a power-law $E^{\gamma}$ to the differential intensity $j(E)$ between 10 and 324 keV. 
    The bottom row gives the speed of the background solar wind.
    The dashed vertical lines indicate the longitudes for which the energy spectra is shown in Figure~\ref{fig:spec_1au}.
    } \label{fig:lon_inner}
\end{figure}

In this section we study the longitudinal variations in the suprathermal particle distributions at different radial distances more closely.  
Zero longitude is along the positive $x$-axis in Figure~\ref{fig:MDI} and the longitude increases anticlockwise.
The top rows of Figure~\ref{fig:lon_inner} show the the omnidirectional intensities as a function of longitude in the solar equatorial plane for simulations without (solid curves) and with (dashed curves) cross-field diffusion. 
The left and right columns show the results at heliospheric radial distances of $r =0.4$~au  and  $r =1.0$~au, respectively. 
Despite that particles are uniformly injected at 0.1~au, the omnidirectional intensities at 0.4~au show a clear longitudinal dependence, as a result of the structured background solar wind where particles propagate. 
Moreover, for the case of no cross field diffusion (the solid curves), when particle's energy is above 500~keV (i.e., above the roll-over energy of $j_{\rm inj}$), the intensities show several `spikes' (near longitudes 0$^{\circ}$, 30$^{\circ}$, 65$^{\circ}$, and 90$^{\circ}$) where the intensities jump by several orders of magnitude.
These spikes correspond to the locations that are magnetically connected to the MIRs (see Figs.~\ref{fig:MDI}b and~\ref{fig:EPs_in_eqplane}d).
For energies higher than $1$~MeV, the  longitudinal width of these intensity spikes at  1 au (right panel of Figure~\ref{fig:lon_inner}) is ${\sim}9^ \circ$, corresponding  to a time-period of ${\sim} 16$~h as seen from the Earth.
As discussed in the previous section, cross-field diffusion will quench the acceleration efficiency near the MIR structures, which explains why the intensity spikes are absent for $a^2 =1/3$ in Figure~\ref{fig:EPs_in_eqplane}).

Apart from the intensity spikes, we see that at 0.4~au the particle intensities show typical variations of a factor of $\sim 6$ within a single energy channel for the simulations without cross-field diffusion. 
Including cross-field diffusion reduces this variation to a factor of ${\sim}3$. 
At 1.0~au, the variation of intensity for  any one of the low energy channels ($< 324$~keV) can reach a factor of $\sim 50$ when $a^2 = 0$ and this variation drops to $10$ when cross-field diffusion is included ($a^2=1/3$). 
 
The second row of Figure~\ref{fig:lon_inner} shows the spectral index $\gamma$, which was obtained by fitting a power-law $E^{-\gamma}$ to the differential intensity $j(E)$ between $10$ and $324$~keV. 
Both simulations without and with cross-field diffusion show an energy spectrum that is harder than the injection spectrum ($j_{\rm inj} \propto E^{-2.5}$). 
Some of this spectral hardening is also seen in simulations with nominal solar wind conditions \citep[e.g,][]{wijsen2020}, where it can be attributed to transport effects.
However, in the current simulations, the variation in the spectral index is strongly influenced by the presence of the CIR structures. 
For example, the CIRs forming ahead of the ``large HSSs'' (e.g., the one covering longitudes -130$^\circ$ to -110$^\circ$0 at 1.0 au) show signatures of local particle acceleration. 
This can be seen by the formation of a double peaked structures in, e.g., the energy channel 136-210 keV in the right panel ($r=1.0$ au). 
This double peak structure is the result of the forward and reverse compression waves both accelerating particles and is often observed at CIRs \citep[e.g.,][]{wijsen2021}. 
The intensity `dip' between the peaks corresponds to the CIR's stream interface (SI), where the compressed and shocked fast wind meets the compressed/shocked slow wind. 

Note that there is a significant difference between the qualitative behaviour of the spectral index at 1~au, depending on whether cross-field diffusion is included or not. 
Without cross-field diffusion the hardest spectra are obtained just upstream the forward and reverse shock waves, and the softest spectra are obtained at the SIs and in the rarefaction regions trailing the HSSs. 
In contrast, when cross-field diffusion is included, the rarefaction regions show the hardest spectra. These results can be understood as the combination of cross-field diffusion reducing the particle acceleration efficiency near the CIR shock waves and being very efficient at spreading particles through rarefaction regions \citep[see also][]{wijsen2019b}.
The SIs remain the regions of the softest energy spectrum, both for the simulations without and with cross-field diffusion.

\begin{figure}[ht]
    \centering
   \includegraphics[width=\textwidth]{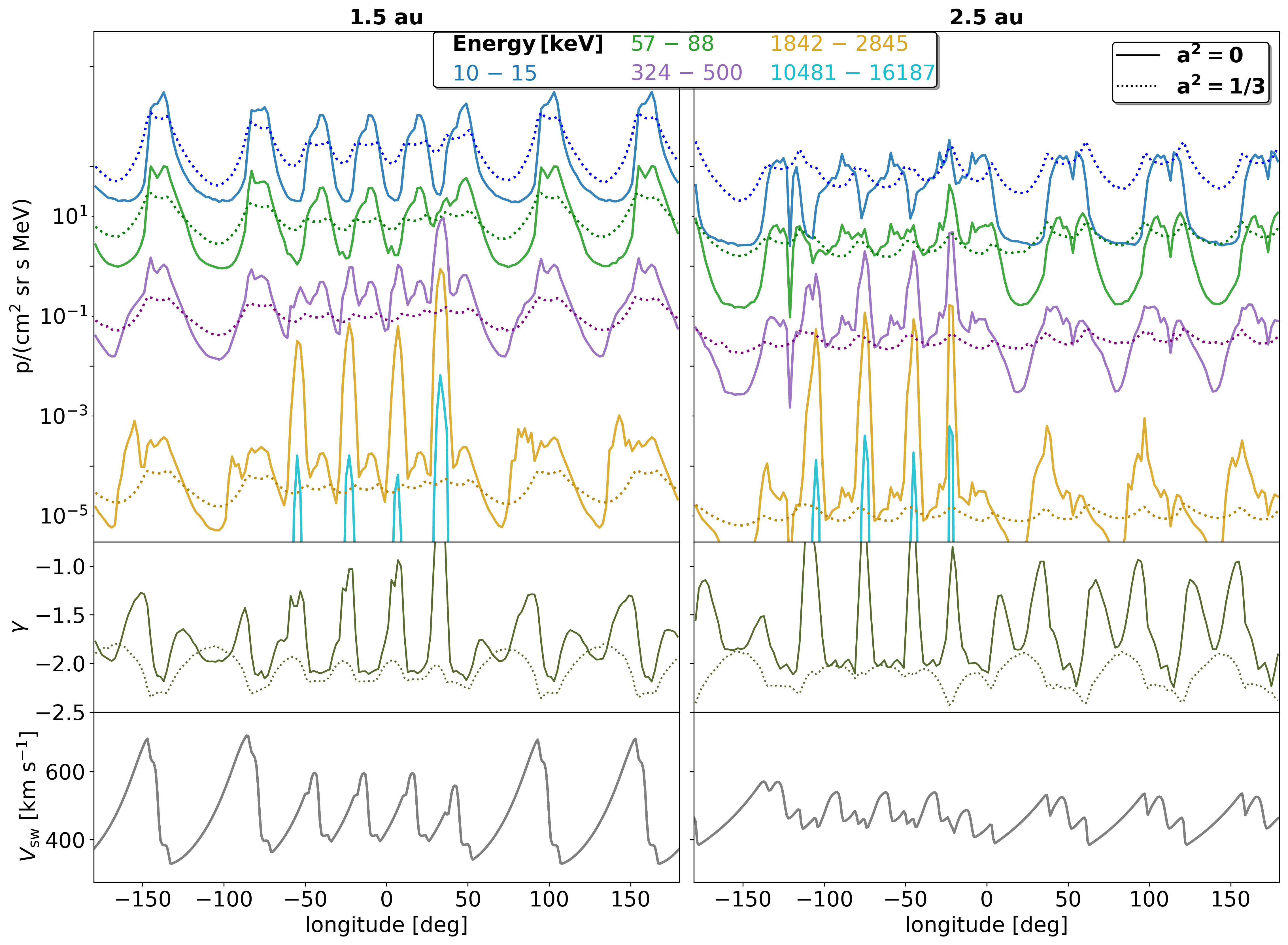}
    \caption{Same as Fig.~\ref{fig:lon_inner}, but for $r=1.5$~au (left) and $r=2.5$~au (right). The number of energy channels in the top row has been reduced for clarity.} \label{fig:lon_outer}
\end{figure}

Figure~\ref{fig:lon_outer} is the same as Figure~\ref{fig:lon_inner}, but for $r=1.5$~au and $r=2.5$~au. For clarity, intensities for only five energy channels are shown.
For the simulation without cross field diffusion (solid curves), the omnidirectional particle intensities are seen to be strongly non-uniform, with variations of more than two orders of magnitude in the lower energy channels. In the highest energy channels,
several clear intensity spikes can again be seen which can be attributed to the MCMT. 
For the simulation with cross-field diffusion, like the cases for $r=0.4$~au and $1.0$ au, the particle intensities show less longitudinal variations. Nevertheless, at 1.5~au the intensity variation can still span more than one order of magnitude for $a^2 = 1/3$ in the lower energy channels. At higher energies, the intensities are almost constant and only vary slightly near very large HSSs.

\subsection{Particle Energy Spectra}

\begin{figure}[ht]
    \centering
    \includegraphics[width=0.99\textwidth]{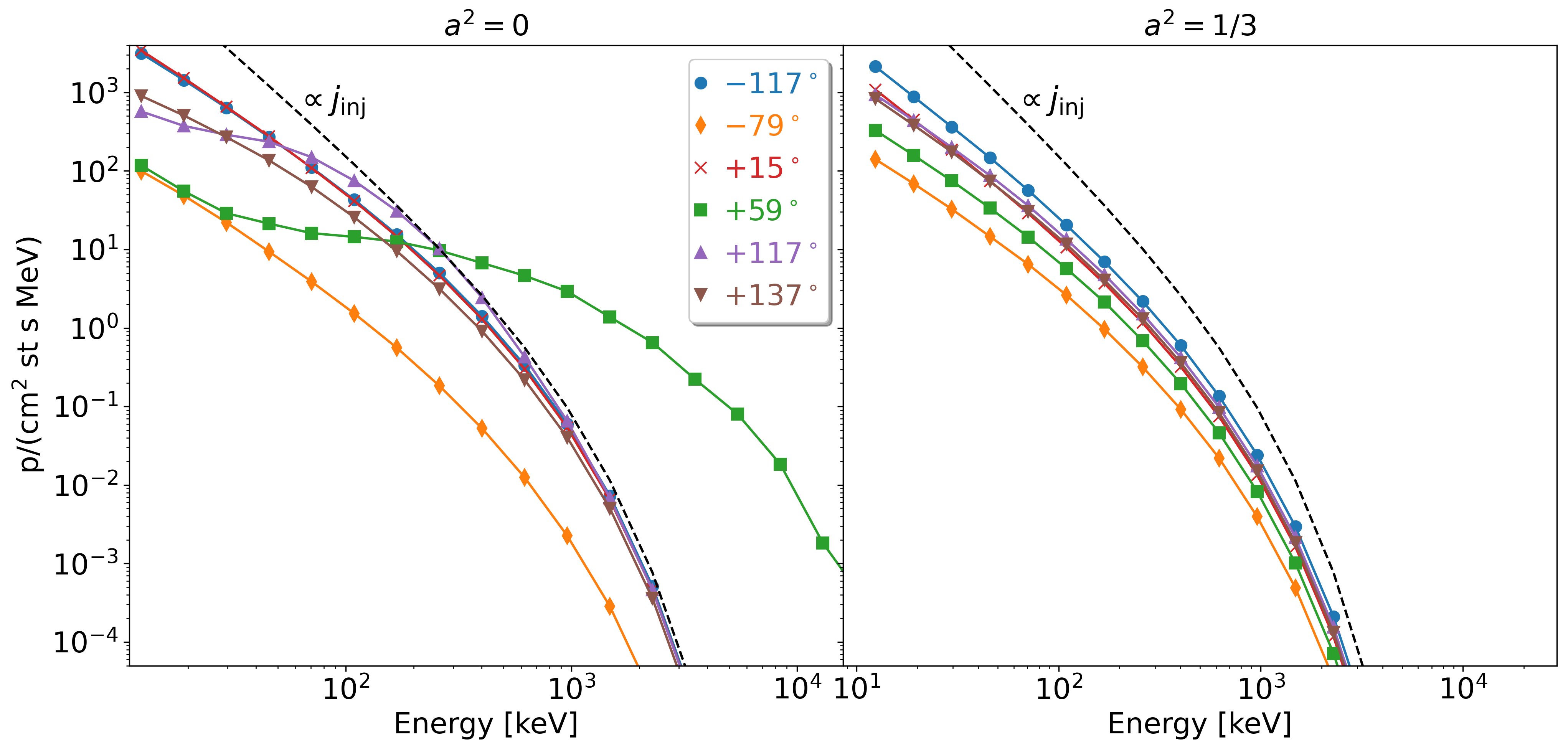}
    \caption{Energy spectra for the different indicated longitudes at 1~au in the solar equatorial plane (see also the vertical dashed lines in the right column of Fig.~\ref{fig:lon_inner}). The left and right panels are for simulations without ($a^2=0$) and with ($a^2=1/3$) cross-field diffusion, respectively. 
    The dashed lines are proportional to the injection spectrum 
    $j_{\rm inj}$.} 
    \label{fig:spec_1au}
\end{figure}

To further demonstrate the difference between the simulations without and with cross-field diffusion, we show in Figure~\ref{fig:spec_1au} the 1~au energy spectra at different longitudes (the selected longitudes are indicated by the vertical dashed lines in the right panel of Figure~\ref{fig:lon_inner}). 
For the simulations with cross-field diffusion (right panel), these spectra do not differ much. 
The similarity of the shape of these spectra can also be seen from the spectral index in the second row of Figure~\ref{fig:lon_inner}.
In contrast, for the simulations without cross-field diffusion (left panel), the energy spectra show large variations. The following features can be deduced:
\begin{enumerate}
    \item The energy spectra of longitudes $-117^\circ$ and $+15^\circ$  overlap  since these locations correspond to the SIs of two different CIRs,  driven by a large and a small HSS, respectively. 
The spectra for these two longitudes, in particular at low energies, are softer than those at other longitudes because the IMF lines passing through these longitudes do not cross any compression waves where particles can get accelerated.  Note that when cross-field diffusion is included, the spectra at these two longitudes no longer overlap. This is because particles can now sample the 3D structure of the underlying (different) CIRs, as they are no longer restricted to their IMF line. 
    \item Longitude $-79^\circ$ corresponds to an intensity minimum in Figure~\ref{fig:lon_inner}, which is a consequence of the underlying solar wind rarefaction region. 
    \item Longitude $+59^\circ$ is magnetically connected to a region where the forward and reverse shocks of two CIRs cross each other. As discussed previously, these merging shocks lead to a very efficient environment (MCMT) for first order Fermi acceleration, therefore producing a harder spectrum and extending to high energies. 
\item 
% separate both observers; 
%\nicolas{ @Gang: The $117^\circ$  observer is located a bit upstream of the reverse shocks. If you zoom in on the figures, you can see that above 120 keV, the intensity peaks at the observer's location, meaning that the most efficient acceleration location for these particles is at slightly larger radial distances than the observer. In contrast, below 100 kev the curves peak exactly at the reverse shock instead of at the observer, indicating that the local acceleration is dominating the intensity profile.    } \gangli{@Nicolas: I think the peaks are either at the dashed line or slightly to the right == Now, as the longitudes increases from 110 to 140, we see in sequence, upstream RS, RS, downstream RS,  SI, downstream FS, (FS), upstream FS.  I think FS itself may or may not be clear from this plot, since the intensity for all energies did not show a jump --- the forward shock may be formed beyond 1 AU. Local acceleration at the RS can be seen for up to 500 keV. Beyond 775 kev, no clear acc can be seen. We can separate these two locations. For the 137 location, emphasize the forward shock is just forming or even not forming at this radius... For 137 location emphasize local acceleration... }
Longitudes $+117^\circ$  and $+137^\circ$ are just upstream a reverse and forward compression wave, respectively. The local intensity peaks of $\lesssim 500$~keV particles at $+117^{\circ}$ indicate that the reverse wave has already steepened sufficiently to allow for notable particle acceleration. In contrast, no clear intensity peaks can be seen at $+137^{\circ}$, indicating that the forward compression wave is less efficient at accelerating particles.   
Both energy spectra show a hardening at lower energies, which is more significant upstream the reverse shock than the forward compression wave.
%\gangli{maybe take this part out until referee asks  (*** This is partly due to the velocity dispersion in the peak intensities that is seen around $115^\circ$ in Figure~\ref{fig:lon_inner}. That is, the lower energy channels peak at the reverse compression wave due to local acceleration, whereas the higher energy channels peak in the fast wind upstream the reverse compression wave, due to particle acceleration beyond 1~au.  ****)}
\end{enumerate}

We next examine the longitude-averaged particle energy spectra at different radial distances.We emphasize that our synthetic solar wind configuration is to be regarded as containing an ensemble of fast and slow wind streams. So these longitude-averaged spectra correspond to a range of energy spectra that a spacecraft would measure over multiple solar rotations.
\begin{figure}[ht]
    \centering
    \includegraphics[width=0.9\textwidth]{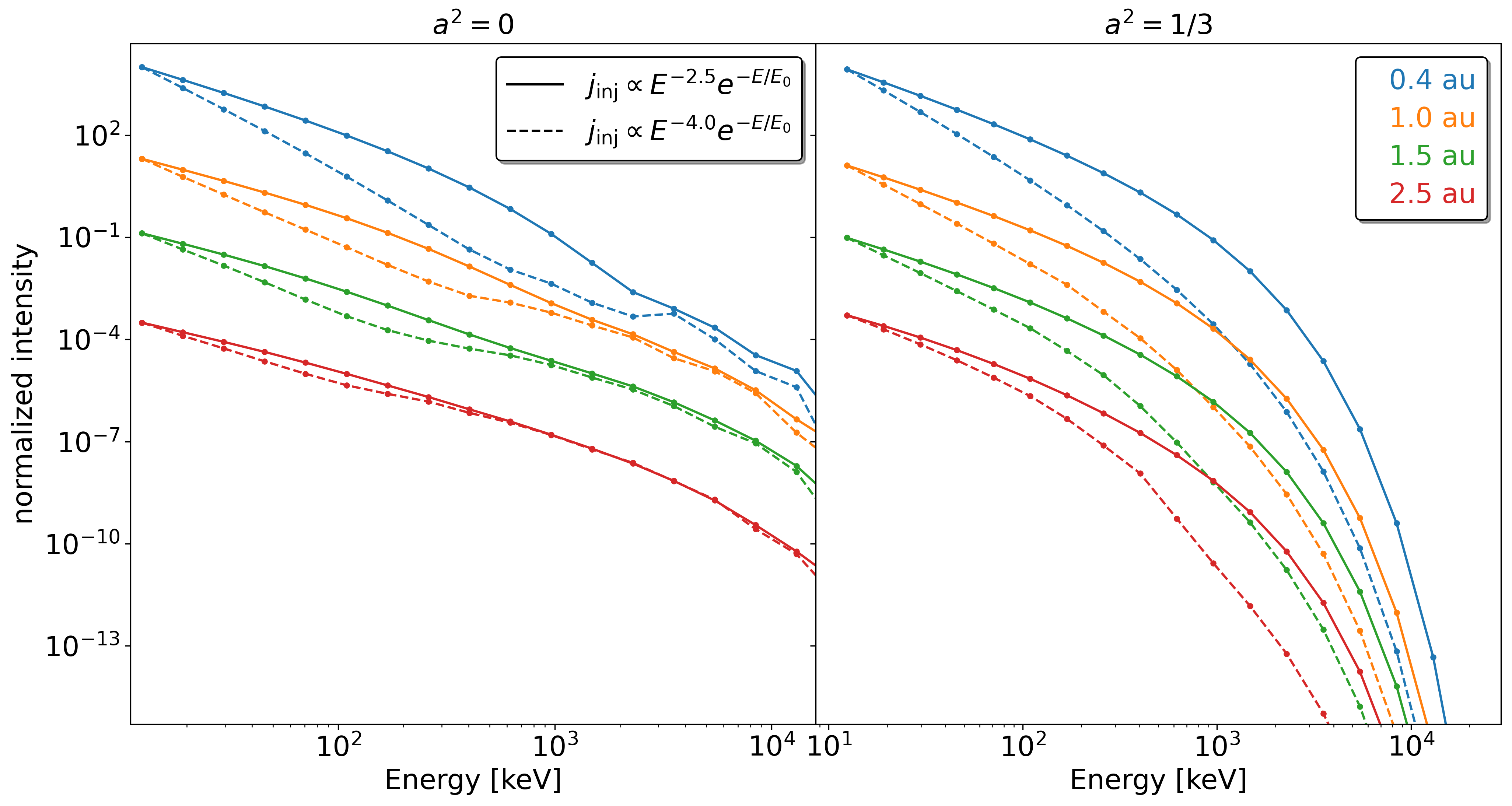}
    \caption{Energy spectra averaged over longitude in the solar equatorial plane at different radial distances. For better comparison, the energy spectra have been normalized to their value at $10$ keV and shifted downward by a factor of $10^{-1.75}$ to avoid overlap.
    } \label{fig:spec_avg}
\end{figure}

Figure~\ref{fig:spec_avg} shows these longitude-averaged spectra at different radial distances. Two different injection spectra are considered: the solid lines are the results for the injection spectrum of the previous sections with  $\gamma=-2.5$, whereas the dotted lines are for a softer injection spectrum with $\gamma=-4$. 
Both injection spectra are assumed to have an exponential roll-over at $500$~keV. 
The energy spectra shown in Figure~\ref{fig:spec_avg} have been normalized to their value at 10~keV and have been shifted downward to avoid overlap and to facilitate their comparison.

The left panel of Figure~\ref{fig:spec_avg} shows that when cross-field diffusion is not included, the longitude-averaged  energy spectra are no longer characterised by a power law with a roll-over at $500$~keV.
That is, towards large radial distances (at $2.5$ au) the energy spectra obtained from the two different injection spectra converge to the same spectral shape. 
At smaller radial distances ($\leq 1$~au), the simulation with the softer injection spectrum (dashed lines) produces a convex-shape like spectrum below $\sim 500$~keV.  
At low energies, the  spectrum resembles that of the injection spectrum, it then hardens before eventually roll over at higher energies. The energy when the hardening starts decreases with radial distance. This behavior is due to two reasons. First, a softer injection spectrum implies that there are relatively more low energy particles and therefore the spectrum at lower energies is more closer to the injection spectrum. Secondly, as explained before,  the acceleration beyond 1~au is more pronounced for the case of no cross-field diffusion. This also helps to lead to the convex-shape shown in the spectra. 
%This is a combined results of a source spectrum at the inner boundary and the modulation of accelerated particles from acceleration sites further away (e.g., $> 2.5$~au).
However, above 500~keV, the energy spectra show similar behaviour for both injection spectra, indicating that this part of the spectra is less dependent on the injection and more a result of the acceleration process.

The right panel of Figure~\ref{fig:spec_avg} is for the case when cross-field diffusion is included. Consider first the case of an injection spectrum with  $\gamma=-2.5$ (solid curves),  the longitude-averaged energy spectra at 
different heliocentric distances are similar. They are all characterised by a power-law with an exponential roll-over. This is because  cross-field diffusion prevents  particles from remaining for an extended period in the regions with ongoing particle acceleration. Therefore, the spectral features are less affected by the interplanetary acceleration processes and resembles more the injection profile.  
In fact,
the roll-over energy are all close to 500 keV, which is the roll-over energy of the injection spectrum at the inner boundary. 
For the case when the injection spectrum is softer ($\gamma=-4.0$ and the dashed curves), we see that the spectra at $0.4$ au and $1.0$ au are again consistent with a power-law and an exponential roll-over. At $r=1.5$ and $2.5$ au, the spectra between $300$ and $800$ keV show some features which are presumably due to the fact that observers are now closer to CIR shocks. Note that the presence of cross-field diffusion can effectively attenuate the acceleration processes at the CIR shocks, which is clearly seen by comparing the right panel to the left panel. 
%Moreover, the energy spectra of the simulation with the softer energy spectra remain significantly softer throughout the entire simulation domain. That is, the solar wind structures only weakly influence the particle populations. 

\section{Summary and Conclusions}\label{sec:conclusions}

In this work we examine how interplanetary conditions may affect suprathermal particle distributions.
Since these suprathermal particles may constitute the seed population of the DSA mechanism at  CME-driven shocks which produce gradual SEP events \citep[e.g.,][]{desai_giacalone_2016}, it is important to understand the processes that can cause variations in this particle population. 
One such process is the existence of velocity shears and speed gradients in the background solar wind.
To study this, we used the PARADISE model to propagate a population of suprathermal particles originated near the Sun (presumably produced continuously at nano and micro flares) through a structured background solar wind, modelled by EUHFORIA that contained several HSSs, rarefaction regions, and CIRs. 

The simulations show that the presence of velocity shears and speed gradients in the  solar wind can significantly modulate a suprathermal particle population. To focus on the effect of the background solar wind structures, we inject a source of  particles with a uniform intensity at the inner boundary of the model. In our simulation, we consider cases with and without cross-field diffusion. 
The lowest and highest particle intensities were obtained inside solar wind rarefaction regions and CIRs, respectively. 
In the simulations without cross-field diffusion, particle acceleration at the CIR compression waves hardened the energy spectrum of the suprathermal population locally. 
This produced significant longitudinal variations in the  simulated energy spectra.
Moreover, the energy spectra averaged over one solar rotation was significantly hardened compared to the injection spectrum. 
At larger radial distances ($\geq 1.5$~au), the average energy spectra was seen to become independent of the injection spectrum and it was thus completely modulated by the underlying solar wind structures. 
The hardening of the energy spectra at large energies ($>1$~MeV) was partly due to particle acceleration occurring in the regions where the forward and reverse shock waves of two different CIRs cross each other, producing an MIR. 
This converging shock pair forms a collapsing magnetic trap configuration which leads to an efficient first order Fermi acceleration process. 
In the simulations, these efficient acceleration regions reside at large radial distances ($> 2$~au) and have a limited spatial extend.  This is in agreement with previous observational studies of CIR suprathermal particles \citep[see e.g.,][]{Mason2008, Ebert2012, Filwett2017} where $> 1$~MeV/nuc particles are likely accelerated beyond 1.5~au.  
These efficient acceleration regions gave rise to intensity `spikes' inside the fast solar wind streams and their trailing rarefaction regions in the inner heliosphere, all the way to the inner boundary of our model. 
As we discussed earlier, at 1~au 
these intensity spikes can last ${\sim} 16$~h, they therefore 
can serve as possible candidate of a transient seed population for DSA at a CME-driven shock, and hence help to explain the large event variability of gradual SEP events.

The inclusion of cross-field diffusion decreases the efficiency of particle acceleration at the CIRs and the MCMTs. This is because the cross-field diffusion tends to spread particles uniformly through space and the compression waves or magnetic traps where particles can be accelerated have a relatively small spatial extend in the simulation. 
That is, most of the solar wind is in an expanding state, where particles will undergo adiabatic cooling.
In the simulation, the reduction in interplanetary particle acceleration was especially notable near the MIR formation region, where cross-field diffusion efficiently removes particles from the collapsing magnetic bottle. 
We note that the presence and strength of interplanetary cross-field diffusion is still a matter of lively debate \citep[see e.g.,][for a recent discussion]{strauss20} and that cross-field diffusion likely varies depending on the underlying solar wind conditions and IMF structures. 
For example, SIs inside CIRs may act as efficient diffusion barriers due to the presence of non-axisymmetric  turbulence \citep[][]{strauss16}. 
Our study illustrates that the properties of a suprathermal particle population will strongly depend on the magnitude of the cross-field diffusion.  A weak cross-field diffusion conditions near solar wind compression waves will lead to an overall hardening of the average energy spectra and  produce strong spatial variations in the local energy spectra of the suprathermal population as well. 
The latter may contribute to the observed strong variability of energetic particle production at CME-driven shocks.
%in their ability of producing long-lasting gradual SEP events.

The EUHFORIA simulation presented in this study did not contain HCS nor CMEs. cMEs  can occur frequently during solar maximum. 
As CMEs typically drive a compression or shock wave, it is to be expected that they will also affect the suprathermal particle population. 
In addition, when a CME takes over a CIR or a preceding CME, a collapsing magnetic trap will form in a similar fashion as the formation of MIR discussed in this paper. Therefore we expect CME-CME and CME-CIR interactions will also play a role in producing seed populations.  
We leave it for a future study to investigate the evolution of the suprathermal particle population in  such solar wind conditions in more detail. 

\section{Open Research}
No input data is used in this simulation work. All simulation result data that lead to the figures produced in this paper are archived in a public dateset on Zenodo. The doi is
\url{https://doi.org/10.5281/zenodo.7672340}.

\acknowledgments
N.W.\ acknowledges acknowledges support from NASA program NNH17ZDA001N-LWS and from the Research Foundation - Flanders (FWO-Vlaanderen, fellowship no.\ 1184319N).
This project has received funding from the European Union’s Horizon 2020 research
and innovation programs under grant agreement No.\ 870405 (EUHFORIA 2.0). These results were also obtained in the framework of the ESA project ``Heliospheric modelling techniques'' (Contract No.\ 4000133080/20/NL/CRS).
Computational resources and services used in this work were provided by the VSC (Flemish Supercomputer Centre), funded by the FWO and the Flemish Government-Department EWI. 
 This work is supported in part by NASA grants 80NSSC19K0075, 80NSSC19K0079, and 80NSSC20K1783 at UAH.
 D.L. acknowledges support from NASA Living
With a Star (LWS) programs NNH17ZDA001N-LWS and
NNH19ZDA001N-LWS, the Goddard Space Flight Center
Internal Scientist Funding Model (competitive work package)
program and the Heliophysics Innovation Fund (HIF) program.
 These results were also obtained in the framework of the projects
C14/19/089  (C1 project Internal Funds KU Leuven), G.0D07.19N  (FWO-Vlaanderen), SIDC Data Exploitation (ESA Prodex-12), and Belspo project B2/191/P1/SWiM. 
R.C.A. acknowledges support from NASA grants 80NSSC21K0733 and 80NSSC21K1307. Supports by ISSI and ISSI-BJ through the international team 469 is also acknowledged.

%% ------------------------------------------------------------------------ %%
%% References and Citations

%%%%%%%%%%%%%%%%%%%%%%%%%%%%%%%%%%%%%%%%%%%%%%%
%
% \bibliography{<name of your .bib file>} don't specify the file extension
%
% don't specify bibliographystyle
%%%%%%%%%%%%%%%%%%%%%%%%%%%%%%%%%%%%%%%%%%%%%%%

%\bibliography{ enter your bibtex bibliography filename here }

%\bibliography{sepref}

\end{document}